%% file: main.tex
\documentclass[aps,pre,reprint,superscriptaddress,floatfix]{revtex4-2}

\usepackage[T1]{fontenc}
\usepackage[utf8]{inputenc}
\usepackage{amsmath,amssymb}
\usepackage{newtxtext,newtxmath}
\usepackage[final]{microtype}
\usepackage{bm}
\usepackage{graphicx}
\usepackage{xcolor}
\usepackage{enumitem}
\setlist{nosep,leftmargin=*}
\usepackage[hidelinks]{hyperref}
\usepackage[nameinlink,noabbrev]{cleveref}
\crefname{equation}{Eq.}{Eqs.}
\Crefname{equation}{Eq.}{Eqs.}

\newcommand{\vect}[1]{\bm{#1}}
\newcommand{\dd}{\mathrm{d}}
\newcommand{\Tr}{\operatorname{Tr}}
\newcommand{\pdet}{\operatorname{det}\nolimits'}
\graphicspath{{figures/}}

\newcommand{\budget}{B}
\newcommand{\Gmat}{\mathsf{G}}
\newcommand{\cpty}{\mathcal{C}}
\newcommand{\Kmat}{\mathsf{K}}
\newcommand{\Amat}{\mathsf{A}}
\newcommand{\Lmat}{\mathsf{L}}
\newcommand{\Hmat}{\mathsf{H}}
\newcommand{\lnk}{\ln \vect{k}}
\newcommand{\score}{\mathcal{Q}}
\newcommand{\etol}{\varepsilon_{\mathrm{tol}}}
\newcommand{\cont}{\kappa}

\begin{document}

\title{Why do continual physical learners need sleep?}
\hypersetup{pdftitle={Why continual physical learners need sleep}}
\author{Nada Elmeligy}
\affiliation{Department of Physics, Syracuse University, Syracuse, NY 13244 USA}
\author{Nidhi Pashine}
\affiliation{Department of Physics, Syracuse University, Syracuse, NY 13244 USA}
\author{J. M. Schwarz}
\affiliation{Department of Physics, Syracuse University, Syracuse, NY 13244 USA}
\affiliation{Indian Creek Farm, Ithaca, NY 14850 USA}

\begin{abstract}
Sleep is widespread across the animal kingdom, including in organisms without a centralized nervous system. Does learning create a need for sleep, even without biology? To explore this question, we study a disordered spring network trained to perform a sequence of mechanical allosteric tasks through local changes in bond stiffness. We find that continual learning progressively reduces the network's ability to acquire new tasks as individual bonds become increasingly specialized. We identify an exact algebraic relationship between the polarization of bond dominance and the network's remaining learning capacity, providing a quantitative measure of this loss of adaptability. We then demonstrate that random internal perturbations allow individual bonds to estimate their mechanical dominance and locally adjust their stiffnesses to restore learning capacity. However, the forces associated with active learning bias these local measurements and interfere with restoration. Separating learning and repair into wake-like and sleep-like phases removes this interference and improves continual-learning performance. Our results suggest that the need for offline restorative phases can emerge from the mechanics of physical learning, even in systems without biological components.
\end{abstract}

\maketitle
\raggedbottom

\section{Introduction}

What is sleep? At its simplest, sleep is a period of rest during which an organism disengages from its surroundings to greater or lesser extents. More precisely, sleep is identified as a reversible state of behavioral quiescence characterized by reduced responsiveness to external stimuli and an increased need for sleep following deprivation~\cite{Campbell1984,Shaw2003}. Historically, sleep research focused primarily on animals with brains, relying on electrophysiological signatures to identify sleep states~\cite{loomis1937cerebral, schulz2022history}. However, Campbell and Tobler questioned the reliance on these signatures, arguing that sleep in other animals should initially be identified through behavior~\cite{Campbell1984}. Such behavioral criteria have since supported the identification of sleep universally across the animal kingdom, from mammals and birds to fish and nematodes~\cite{Campbell1984,Lakhiani2023,Shaw2000,Raizen2008}. Sleep has even been identified in animals without a centralized nervous system, including jellyfish, hydra and sea anemones~\cite{Nath2017,Kanaya2020,Aguillon2026}. 

The widespread occurrence of sleep raises the possibility that its underlying functions extend beyond the particular biological mechanisms through which it is implemented. After all, living organisms are also physical systems that interact with their environments and adapt by modifying their internal structure and dynamics. Yet adaptation can come at a cost. As a system learns from its interactions with the environment, the resulting changes in its internal structure and dynamics may reduce its ability to adapt to new experiences~\cite{Tononi2014}. Continued learning may therefore require periods of reorganization to restore this capacity for further adaptation~\cite{Tononi2014}. This raises a fundamental question: {\it Does learning create a need for sleep, even without biology?}

The possibility that learning itself creates a need for sleep has also been explored in artificial intelligence. Beginning with the wake–sleep algorithm for training generative neural networks~\cite{Hinton1995}, subsequent studies have demonstrated that sleep-like phases involving offline replay or internally generated activity can mitigate catastrophic forgetting and support continual learning~\cite{Kemker2018,Golden2022,Tadros2022}. These results establish that sleep-like phases can benefit learning systems without biological components. However, conventional artificial neural networks rely on mathematically prescribed learning algorithms. In biological systems, learning is ultimately implemented through physical processes that modify how the system responds to its environment. This raises a related question: Can simpler, nonbiological physical systems, whose own interactions and dynamics provide the mechanisms for learning, also develop a need for sleep-like phases?

 Physical learning offers a useful framework to begin to answer this question. In physical learning, the interactions and dynamics of a material serve as the mechanisms for learning, allowing the material to be trained to perform prescribed tasks without relying on conventional computational learning algorithms~\cite{SternMurugan2023,anisetti2023emergent,momeni2025training}. For example, a material like a disordered elastic network can be trained to acquire new functions not through a centralized processor, but by applying local update rules directly to the network's own structure~\cite{Scellier2017,Stern2021,pashine2021local,Anisetti2023,Anisetti2024}. Here, the trained function will be mechanical allostery between an input bond and an output bond: when a force dipole is applied across the input bond, the output bond should respond with a prescribed tension~\cite{Rocks2017,pashine2021local}. We achieve this by explicitly updating the trainable degrees of freedom, namely, the individual bond stiffnesses. Because these networks are nonbiological, they allow us to investigate whether the need for offline restorative phases can arise from the mechanics of continual learning.

\begin{figure*}[t]
  \centering
  \includegraphics[width=\textwidth]{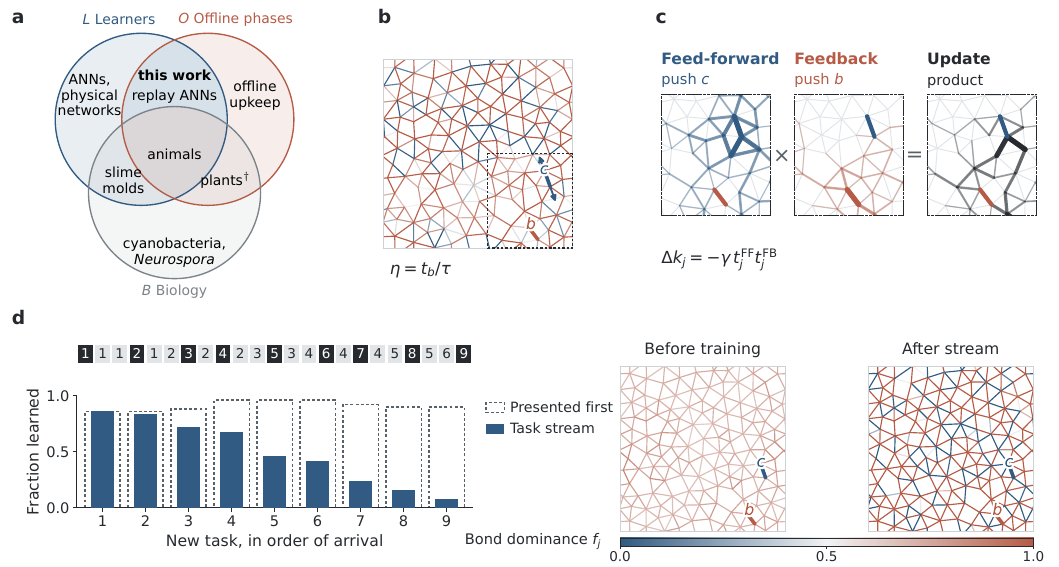}
  \caption{\textbf{Continual training reduces new-task learning in an elastic network.}
(a)~Conceptual map of biology ($B$), learners ($L$), and offline phases ($O$). The highlighted region, $L\cap O\setminus B$, contains this work and artificial networks with sleep-like replay~\cite{Tadros2022}. Examples illustrate the sets; $^\dagger$plant learning and the identification of nyctinasty with sleep are debated.
(b)~The network shown has 128 nodes and 355 bonds. A task applies input tension $\tau$ at port $c$ (blue) and measures $\eta=t_b/\tau$ at port $b$ (red). Other bond colors show bond dominance $f_j=k_j/(k_j+k_{\mathrm{env},j})$, using the shared scale in (d). The dashed window is enlarged in (c).
(c)~The first wake step of the stream: line widths show the magnitudes of feedforward-state tension, error-driven feedback-state tension, and their product on trainable bonds. The equation gives the signed local stiffness update.
(d)~Wake-only training over 23 visits, each with 400 steps, following a three-task prelude. Stream tasks are numbered 1--9 by first arrival; dark tiles mark first visits and light tiles mark repeats. Filled bars give the fraction of 50 networks learning each new task at its first visit, with $|\eta(400)+0.4|\leq0.10$. Dashed outlines give controls in which each task is instead trained first from the same post-prelude state. Maps show the bond dominance for one realization before any training and after the prelude and full stream.}
  \label{fig:setup}
\end{figure*}

Here, we demonstrate that continual learning in a disordered spring network can reduce its capacity to acquire new tasks, creating a need for restorative changes to its mechanical properties (Fig. 1). We identify a quantitative relationship between the specialization of individual bonds and the network's remaining learning capacity. We then show that local measurements of mechanical fluctuations can guide restorative updates that recover this capacity. However, ongoing task activity interferes with these measurements, providing a physical reason to separate learning and repair into wake-like and sleep-like phases. Our results suggest that the need for offline restoration can arise from the mechanics of continual learning, even in the absence of biology. Finally, we discuss how these findings may inform our understanding of sleep and learning in biological systems, including the role of sleep in restoring neural plasticity.

\section{Theory}
\label{sec:model}

\subsection{Network and bond dominance}
We consider linear networks with positive bond coefficients and a fixed compatibility matrix $G$, whose response is governed by the quadratic form below. This structure is shared by spring networks, resistor networks, and linear flow networks: the coefficients are stiffnesses, electrical conductances, or hydraulic conductances, respectively. The same derivation applies to artificial linear models admitting this structure. We will use notation for mechanical networks throughout. For springs, the quadratic form is the stored elastic energy; for resistive and flow networks, it is a dissipation potential,
\begin{equation}
  E = \sum_{b} \tfrac{1}{2}\, k_b\, e_b^2 ,
  \qquad
  \vect{e} = \Gmat\,\vect{u}. 
  \label{eq:energy}
\end{equation}
Note that $\vect{u}$ denotes the vector of node displacements and $\vect{e}$ the vector of bond extensions. In resistor and flow networks, $\vect{u}$ instead denotes node voltages or pressures, and $\vect{e}$ their differences across bonds. Each row of
$\Gmat$, denoted by $\mathbf{g}_b^{T}$, specifies the extension
of bond $b$ in terms of the nodal displacements, so that
$e_b = \mathbf{g}_b^{T}\mathbf{u}$.

The mechanical response of the network is determined by the
Hessian $\Hmat$, obtained by taking the second derivatives of
the elastic energy with respect to the nodal displacements.
Writing $\Kmat=\mathrm{diag}(k_b)$ for the diagonal matrix of
bond stiffnesses, we obtain
\begin{equation}
\Hmat = \Gmat^T \Kmat \Gmat.  
 \label{eq:hessian}
\end{equation}
$\Hmat^{+}$ denotes its pseudo-inverse on the allowed node subspace with zero modes removed. 

To characterize how individual bonds contribute to the
mechanical response of the network, we introduce the
bond dominance $f_b$. Consider applying a force dipole
across the endpoints of bond $b$. After the network
relaxes, the applied load is distributed between the
bond itself and the surrounding network. We define
$f_b$ as the fraction of this load carried by bond $b$,
\begin{equation}
f_b \equiv k_b \mathbf{g}_b^T \Hmat^+ \mathbf{g}_b.
\end{equation}
 By construction,
$0 \leq f_b \leq 1$. The dominance can also be expressed in terms of the
effective stiffness $k_{\mathrm{env},b}$ of the
surrounding network between the endpoints of bond $b$:
\begin{equation}
f_b = \frac{k_b}{k_b+k_{\mathrm{env},b}}
    = \frac{S_b}{1+S_b},
\qquad
S_b \equiv k_b \mathbf{g}_b^T \Hmat_{-b}^+
\mathbf{g}_b.
\label{eq:loadfrac}
\end{equation}
Here, $\Hmat_{-b}$ is the Hessian with bond $b$ removed.
The cavity expression follows from the
Sherman--Morrison formula~\cite{ShermanMorrison1950}, provided that
removing the bond preserves the rank of the Hessian. Please note that $S_b=k_b/k_{\mathrm{env},b}$ is the dimensionless relative stiffness of the network with the bond $b$ removed. In other words, it measures the stiffness of bond $b$ relative to the effective stiffness of the surrounding network.

 Given this construction, when an external load is applied across the two nodes joined by bond $b$, the network redistributes it as it relaxes. The dominance $f_b\in[0,1]$ is the share of this load that stays in bond $b$, while the rest of the network carries the remaining share $1-f_b$ (which sets the local modulus introduced by Hexner et al.~\cite{Hexner2018}). This load can be a dipole force, a current, or a flow in the respective systems. A bond with no alternative load paths has $f_b=1$. To understand how the dominances of individual bonds are related to one another, we introduce a projection operator in bond space. Writing the Hessian as $\Hmat=M^TM$, where $M=\Kmat^{1/2}\Gmat$, we recognize that the columns of $M$ span the space of stiffness-weighted compatible bond extensions. The orthogonal projector onto this space is therefore
\begin{equation}
\Amat \equiv \Kmat^{1/2}\, \Gmat\, \Hmat^{+}\, \Gmat^{\mathsf{T}}\, \Kmat^{1/2},
  \qquad \Amat^2 = \Amat ,
  \qquad \Amat_{bb} = f_b ,
  \label{eq:projector}
\end{equation}
which projects onto the space of stiffness-weighted
compatible bond extensions. Its complement, $I-A$,
projects onto the corresponding space of
stiffness-weighted states of self-stress~\cite{Calladine1978}.
Here, $N_b$ denotes the number of bonds,
$n=\mathrm{rank}(\Hmat)$ is the number of independent
deformation modes, and $R=N_b-n$ is the number
of independent states of self-stress. As we show below, the projector structure imposes constraints on how dominance can be redistributed among bonds as their stiffnesses change.

\subsection{Task response}

We consider allosteric tasks in which an applied force
dipole at one bond produces a prescribed tension at
another bond. Each task $\alpha=(p,q)$ consists of
an input bond $p$ and an output bond $q$, whose
stiffnesses are held fixed during training. Applying a force dipole of magnitude $\tau$ across
the endpoints of bond $p$ produces a nodal force
$\mathbf{F}=\tau\mathbf{g}_p$ and a displacement
field $\mathbf{u}=\tau H^+\mathbf{g}_p$.
The resulting tension in bond $q$ is
$t_q=k_q\mathbf{g}_q^T\mathbf{u}$.
The normalized task response is therefore
\begin{equation}
\eta_\alpha \equiv \frac{t_q}{\tau}
= k_q\mathbf{g}_q^T \Hmat^+\mathbf{g}_p
= \sqrt{\frac{k_q}{k_p}}A_{qp}.
 \label{eq:task}
\end{equation}
Since the port stiffnesses are held fixed,
training modifies the task response through changes
in the coupling $A_{qp}$ between the two bonds.

\subsection{Plasticity bound}
\label{theory:load-response}

Learning adjusts the bond coefficients and can thereby redistribute the dominances $f_b$, changing how the network carries an applied load. To quantify this redistribution, we consider the response of the bond dominances to small relative changes in stiffness. The changes are governed by the bond Laplacian $\Lmat$ (Supplemental Material, Sec.~\ref{app:laplacian}):
\begin{equation}
  \delta \vect{f} = \Lmat\, \delta \lnk ,
  \qquad
  \Lmat_{bb} = f_b(1-f_b),
  \qquad
  \Lmat_{bc} = -\Amat_{bc}^2 .
  \label{eq:laplacian}
\end{equation}
The diagonal elements of $\Lmat$ measure the sensitivity of a
bond's dominance to changes in its own stiffness, while the
off-diagonal elements describe how such changes redistribute
dominance among other bonds. Because $\Amat$ is a projection
operator, $\Lmat$ has zero row sums and is positive semidefinite.

The sensitivity of the task coupling $\Amat_{qp}$ to the stiffness of a trainable bond $b$ is bounded (Supplemental Material, Sec.~\ref{app:landing}):
\begin{equation}
  \left| \frac{\partial \Amat_{qp}}{\partial \ln k_b} \right|
  = \left| \Amat_{qb}\,\Amat_{bp} \right|
  \le \tfrac12\, f_b\,(1-f_b) .
  \label{eq:plasticity}
\end{equation}
We therefore define $f_b\,(1-f_b)$ as the \emph{bond plasticity}. It limits the change in a task response produced by a small relative change in that bond's stiffness. Bond plasticity vanishes as $f_b$ approaches either zero or one, indicating that both highly dominant
and nearly bypassed bonds have limited capacity to
contribute to further learning.

\subsection{A conserved learning capacity-polarization budget}
\label{theory:budget}

At fixed topology, boundary conditions, and rank, the sum of the bond dominances is conserved (Supplemental Material, Sec.~\ref{app:budget}). Since the dominances are the diagonal elements of the projection operator, we have 
\begin{equation}
\sum_b f_b = \mathrm{Tr}\,\Amat =\Tr\!\left( \Hmat^{+}\Hmat \right) = n, 
  \label{eq:sumrule}
\end{equation}
where we have used the cyclic invariance of the trace. Also, although the Hessian $\Hmat$ acts on nodal displacements and the projector $\Amat$ acts on bond extensions, their ranks are equal, with the trace equaling the rank~\cite{HoaglinWelsch1978}; for resistor networks, \cref{eq:sumrule} reduces to Foster's theorem~\cite{Foster1949}. Because $\sum_b f_b$ is fixed, training cannot create dominance, only reallocate it between coupled bonds [\cref{eq:laplacian}]. Since $\sum_{c\neq b}\Amat_{bc}^2 = f_b(1-f_b)$, this coupling vanishes at both extremes: a fully dominant bond ($f_b=1$) is mechanically isolated from the rest of the network, while a bypassed bond ($f_b\to 0$) carries no load to redistribute.

We define the network's learning capacity as the sum of the individual bond plasticities, and the polarization (or sleep pressure) as the summed squared deviation of the dominances from their mean,
\begin{equation}
  \cpty \equiv \sum_b f_b\,(1-f_b),
  \qquad
  P \equiv \sum_b \left( f_b - \bar{f} \right)^2 ,
  \qquad \bar f = \frac{n}{N_b},
  \label{eq:capacity}
\end{equation}
and \cref{eq:sumrule} ties them together. We thus identify an exact constraint relating the network's learning capacity and dominance polarization (Supplemental Material, Sec.~\ref{app:budget}):
\begin{equation}
  \boxed{\;\cpty + P = \frac{n R}{N_b} \equiv \budget\;}
  \label{eq:budget}
\end{equation}

This identity is independent of the learning rule and of whether the network is mechanical, electrical, or hydraulic. The conserved quantity is the budget $\budget$: learning capacity and polarization can change, with $\Delta\cpty=-\Delta P$. Thus, any learning-induced increase in polarization spends learning capacity, while reducing polarization restores it. Whether training increases polarization depends on the learning dynamics; the conservation law determines the accompanying change in learning capacity. In the Results section, we investigate whether continual training increases dominance polarization and thereby reduces the learning capacity of a disordered spring network.

\subsection{Repair}
\label{theory:repair}

Having established that increased dominance polarization reduces learning capacity, we now ask whether changes in bond stiffness can reverse this process. We define repair as stiffness adjustments that reduce polarization and thereby restore learning capacity $\cpty$. Physically, dominance $f_b = S_b/(1+S_b)$ sets the mechanical balance between bond $b$ and the surrounding network: a bond with $f_b \to 1$ dictates local deformation without yielding to its neighbors, whereas a bond with $f_b \to 0$ yields passively without transmitting force. A single bond in isolation has maximum plasticity $f_b(1-f_b)$ when it is evenly matched to its surroundings ($f_b = 1/2$). Across the network, however, all bonds compete for the fixed total dominance $\sum_b f_b = n$ [\cref{eq:sumrule}]. Because one bond can move closer to dominance only by pushing others toward passivity, the network maximizes its collective learning capacity $\cpty$, and minimizes polarization $P$ [\cref{eq:budget}], when every bond claims an equal share of the total dominance, $f_b = \bar{f} = n/N_b$.

To construct a repair rule, we use \cref{eq:loadfrac}, which shows that increasing a bond's stiffness increases its own dominance, while decreasing its stiffness has the opposite effect. This suggests a feedback mechanism in which bonds with above-average dominance are weakened and those with below-average dominance are strengthened. We choose the rate of adjustment to be proportional to the deviation $f_b-\bar f$. Since \cref{eq:laplacian} relates changes in dominance to changes in logarithmic stiffness, we formulate the rule in terms of $\ln k_b$, corresponding to relative rather than absolute stiffness changes and preserving positive stiffnesses. Assuming all bond stiffnesses are free to adjust, the repair rule is
\begin{equation}
  \frac{\dd \ln k_b}{\dd t} = -\varepsilon \left( f_b - \bar f \right).
  \label{eq:creepmean}
\end{equation}
Here $\varepsilon>0$ sets the repair rate, with time measured in units of one update of all bonds. With $\vect{q}=\vect{f}-\bar f\,\vect{1}$, the dominance response in \cref{eq:laplacian} gives
\begin{equation}
  \dot P = -2\varepsilon\,\vect{q}^{\mathsf T}\Lmat\vect{q} \leq 0,
  \qquad \dot{\cpty}=-\dot P\geq 0,
  \label{eq:repair-descent}
\end{equation}
because $\Lmat$ is positive semi-definite (Supplemental Material, Sec.~\ref{app:laplacian}). Equation~\eqref{eq:repair-descent} concerns the ideal continuous-time rule on all bonds. However, the simulations fix the task ports and bound the finite updates, as specified in Supplemental Sec.~\ref{sm:simulation}.

\begin{figure}[!t]
  \centering
  \includegraphics[width=\columnwidth]{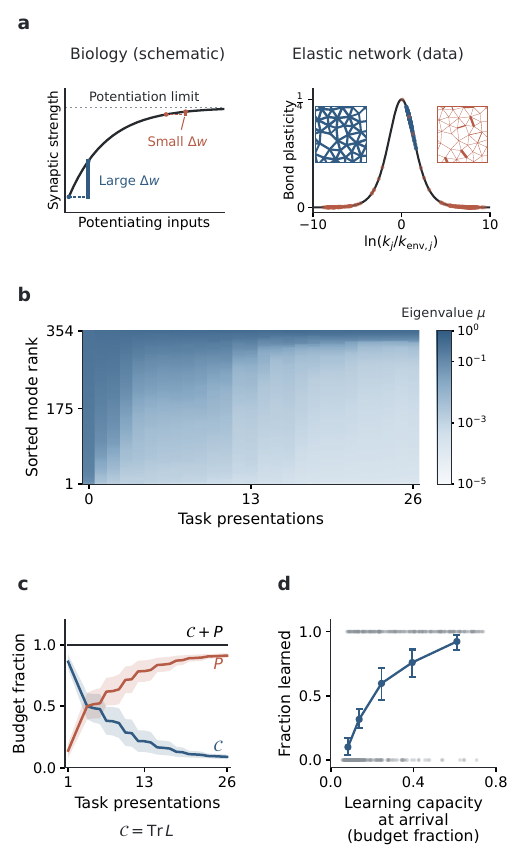}
  \caption{\textbf{Continual learning produces a mechanical analogue of synaptic saturation.}
(a)~Schematic of synaptic saturation~\cite{Tononi2014,Kuhn2016} and bond plasticity $f_j(1-f_j)$ before training (blue) and after the stream (red); inset bond thickness encodes bond plasticity.
(b)~Sorted nonzero eigenvalues of $\Lmat$ through training; color is logarithmic.
(c)~Learning capacity, pressure, and their conserved sum, normalized by $\budget$; means and standard deviations over 50 networks.
(d)~New-task success versus learning capacity at arrival. Gray: 450 individual outcomes; blue: binned success fractions with 95\% whole-network bootstrap intervals.}
  \label{fig:resource}
\end{figure}

\subsection{Local measurements}
\label{theory:measurement}

The repair rule in Eq. (12) requires each bond to know its dominance $f_b=k_b\mathbf{g}_b^T \Hmat^+\mathbf{g}_b$, which depends on the mechanical response of the entire network. How can a bond obtain this information using only locally accessible quantities? Random perturbations have been shown to reveal information about the learned structure of physical networks~\cite{stern2025physical}. Here, we show that fluctuations generated
throughout the network allow each bond to infer its dominance from measurements of its own extension. To accomplish this, we probe the network with internally generated fluctuations while task drives are off. Each bond generates an independent, zero-mean random force,
\begin{equation}
F_b=\sqrt{\vartheta k_b}\,\xi_b,
\qquad
\langle\xi_b\xi_c\rangle=\delta_{bc}.
\end{equation}
Here, $\vartheta$ sets the fluctuation strength. The scaling of the force amplitude with $\sqrt{k_b}$ is chosen so that the covariance of the resulting nodal forces is proportional to the network Hessian:
\begin{equation*}
\langle\mathbf{F}\mathbf{F}^T\rangle
=\vartheta \Gmat^T \Kmat \Gmat
=\vartheta \Hmat,
\end{equation*}
where $\mathbf{F}=\Gmat^T\mathbf{F}_{\mathrm{bond}}$. Consequently, the displacement covariance is $\langle\mathbf{u}\mathbf{u}^T\rangle=\vartheta \Hmat^+$ on the nonzero-mode subspace. The mean bond energy therefore satisfies
\begin{equation}
\langle k_b e_b^2\rangle=\vartheta f_b.
\label{eq:equip}
\end{equation}
Thus, although bond dominance depends on the inverse Hessian, it can be inferred from the fluctuations of a locally measurable quantity, the bond extension (Supplemental Material, Sec.~A~4).

We can therefore construct a local estimator of dominance,
$\hat f_j=k_j e_j^2/\vartheta$, and substitute it into
Eq. (12) to obtain the repair rule
\begin{equation}
\frac{\dd\ln k_j}{\dd t}
=-\varepsilon\left(\frac{k_j e_j^2}{\vartheta}-\bar f\right).
\label{eq:creep}
\end{equation}
Individual measurements fluctuate, but the estimator is unbiased under the prescribed probing protocol. When stiffness adjustments are slow compared with the probing dynamics, the repair rule recovers \cref{eq:creepmean} on average. Although the fluctuations are generated throughout the network, each bond requires only its own stiffness and extension to implement the update. Finally, because repair redistributes bond stiffnesses without reference to previously learned tasks, it may also modify the responses that encode those tasks (Supplemental Material, Secs.~A~3 and A~5).

\subsection{Task contamination and its sign}
\label{theory:contamination}

During learning, the probe and the task act on the same bonds, and the energy each bond senses includes the strain of the task (Supplemental Material, Sec.~\ref{app:contamination}),
\begin{equation}
  \left\langle \frac{k_b\, e_b^2}{\vartheta} \right\rangle
  = f_b + \cont_b ,
  \qquad
  \cont_b \equiv \frac{k_b \left(e_b^{\mathrm{task}}\right)^2}{\vartheta} \;\ge\; 0 .
  \label{eq:contam}
\end{equation}
The task-dependent term shifts the mean repair update by $-\varepsilon\cont_b$ relative to the undriven case. Since $\cont_b \ge 0$, the energy a bond reads during learning never understates its dominance,
\begin{equation}
  f_b \le \left\langle \frac{k_b\, e_b^2}{\vartheta} \right\rangle .
  \label{eq:sign}
\end{equation}
Consequently, probing during active learning mistakes task-carrying bonds for overly dominant ones and systematically biases repair toward softening the very pathways currently in use. Conversely, because task strain only inflates energy readings, any bond that still reads below $\bar{f}$ during wake is genuinely under-dominant ($f_b \le f_b + \cont_b < \bar{f}$). This one-sided contamination motivates a split protocol: retain positive repair increments during wake and reserve negative increments for an undriven offline phase, while leaving the wake learning rule free to adjust stiffness in either direction.

\section{Results}
\label{sec:results}

In Sec.~\ref{theory:budget} we showed that the conservation law $\cpty + P = \budget$ fixes the trade-off between polarization and plasticity for any network of the form in \cref{eq:energy}, but whether training actually depletes learning capacity depends on how the physical degrees of freedom evolve during task acquisition. We now turn to numerical simulations of disordered spring networks to see how local stiffness adaptation navigates this budget across a continual stream of tasks. We also ask whether restorative repair is needed and whether it should occur in a phase separate from active learning.

To train a network on an allosteric task during wake, we adapt a two-state local learning protocol (Fig.~\ref{fig:setup}c)~\cite{Anisetti2023,Anisetti2024}. In the feedforward state, a force dipole of amplitude $\tau$ acts across input bond $p$, generating a tension $t_q^{\mathrm{FF}}$ at output bond $q$. This generated tension across the output bond measures the current task response. The deviation of the output bond tension, in the feedforward state, from the target tension $t_q^\star = \tau \eta^\star$ sets the feedback amplitude. The feedback is applied in the feedback state: with the input drive removed, a force dipole of amplitude $t_q^\star - t_q^{\mathrm{FF}}$ acts across the output bond. Bond stiffnesses are held fixed between the two states, and each trainable bond $j$ updates its stiffness directly from the product of its local feedforward-state tension $t_j^{\mathrm{FF}}$ and feedback-state tension $t_j^{\mathrm{FB}}$:
\begin{equation}
  \Delta k_j = -\gamma\,t_j^{\mathrm{FF}}t_j^{\mathrm{FB}},
  \label{eq:mml}
\end{equation}
with $\gamma = 10^4$ (Supplemental Sec.~\ref{sm:units}), while geometry, connectivity, and port stiffnesses remain fixed.

We evaluate this learning rule across an ensemble of 50 spring networks (128 nodes each; Supplemental Sec.~\ref{sm:simulation}). After an initial three-task prelude trained without repair, identical copies of each network are subjected to a continual stream of new and repeated tasks under different repair protocols. This paired design allows us to isolate how continual wake training polarizes bond dominance (Sec.~\ref{sec:resource}), how local fluctuation-guided repair restores learning capacity and exhibits compensatory rebound after deprivation (Sec.~\ref{sec:repair}), and why task contamination forces weakening repair into an offline sleep-like phase (Secs.~\ref{sec:separation} and~\ref{sec:sign}).

\subsection{Learning uses up the capacity to learn}
\label{sec:resource}

As the network is presented with new tasks, its ability to learn new ones declines. Across a 23-visit stream introducing nine new tasks interleaved with 14 repeats, a progressively smaller fraction of network realizations successfully learns the later tasks (Fig.~\ref{fig:setup}d). This decline could be attributed either to the network's training history or to differences in task difficulty. We test this by training the network on each of the new tasks starting from the same post-prelude (three training tasks) state before the stream. The same tasks are learned more often from this initial state than during the stream, showing that earlier training indeed reduces the network's adaptability.

Continual training modifies the distribution of bond dominance, as illustrated by the network maps in Fig.~\ref{fig:setup}d. Before any training, dominances cluster near the uniform value $\bar f$. Training polarizes the network. Some bonds come to carry nearly all of the load applied across their endpoints, while others carry very little. This polarization reduces the bond plasticity, $f_b(1-f_b)$ (Fig.~\ref{fig:resource}a). Through the plasticity bound in Sec.~\ref{theory:load-response}, this reduction restricts how strongly small relative stiffness adjustments can alter future task responses.

Summing the individual bond plasticities gives the total learning capacity $\cpty$. The conserved budget $\cpty+P=\budget$ in Eq.~\eqref{eq:budget} locks this total into a direct trade-off with dominance polarization~$P$: as polarization rises through training, total learning capacity drops by exactly the same amount (Fig.~\ref{fig:resource}c). At the network level, this depletion appears as a downward shift in the eigenvalue spectrum of the bond-coupling Laplacian $\Lmat$ (Fig.~\ref{fig:resource}b). Small eigenvalues correspond to directions in stiffness-parameter space along which changes produce little redistribution of dominance. Bhaumik and Hexner observed a similar loss of trainability, accompanied by low-frequency vibrational modes, when increasing the number of simultaneously controlled targets under rest-length training~\cite{BhaumikHexner2022}. Here, the loss develops sequentially across a task stream as stiffness adaptation polarizes dominances at fixed geometry.

The scalar learning capacity $\cpty$ tracks the network's ability to acquire the next allosteric task. We measure $\cpty$ at each new task's arrival and track subsequent training success. Tasks arriving at higher learning capacity are learned far more often (Fig.~\ref{fig:resource}d). This positive association remains after accounting for network identity, task arrival position, and repair protocol (Supplemental Fig.~\ref{fig:s-arrival}). Together, these dynamics provide a mechanical analogue of synaptic saturation under the synaptic homeostasis hypothesis, in which wake potentiation strengthens selected synapses at the cost of limiting further plasticity~\cite{Tononi2014}. Reduced capacity for further plasticity has also been observed in human cortex following sleep deprivation~\cite{Kuhn2016}. In our network, bond plasticity vanishes at both dominance limits, $f_b\to0$ and $f_b\to1$. The accompanying rise in polarization $P$ supplies a direct physical measure of learning-induced sleep pressure, paralleling the link between learning and subsequent local sleep homeostasis~\cite{Huber2004}.

\subsection{Local repair supports continued learning}
\label{sec:repair}
\label{sec:demand}

\begin{figure}[!tb]
  \centering
  \includegraphics[width=\columnwidth]{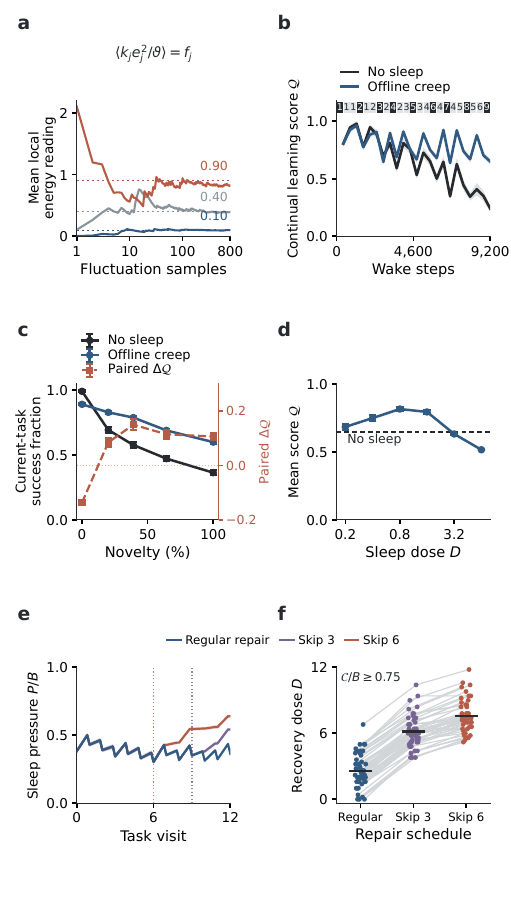}
  \caption{\textbf{Local repair supports continual learning and compensatory recovery after deprivation.}
(a)~Running local-energy averages approach exact dominances (dashed).
(b)~Continual-learning score $\mathcal Q$ across the task stream; strip: task order.
(c)~Current-task success (left) and paired $\Delta\mathcal Q$ (red, right) versus novelty.
(d)~Mean $\mathcal Q$ versus dose; dashed: no sleep.
(e)~Pressure with regular repair or the final three or six bouts omitted; dotted lines mark deprivation onset. Within-visit spacing is schematic.
(f)~Additional recovery dose to reach $\mathcal C/B\geq0.75$; the recovery target is imposed. Dots: 50 paired networks; lines: pairs; bars: means.
Means and standard errors over 50 paired networks in (b--e); (b,d) use $39.1\%$ novelty. Protocols and scoring are given in Supplemental Secs.~\ref{sm:schedules} and~\ref{sm:statistics}.}
  \label{fig:repair}
\end{figure}

To restore the learning capacity lost during wake training, we introduce an offline repair phase after each task visit. With the task drive and learning rule turned off, bonds adjust their stiffnesses to reduce polarization $P$ according to Sec.~\ref{theory:repair}. Although exact calculation of a bond's dominance requires the non-local pseudo-inverse of the network Hessian, probing the quiescent network with local random forces allows each bond to estimate its own dominance locally from its mean elastic energy (Sec.~\ref{theory:measurement}; Fig.~\ref{fig:repair}a). Inserting these fluctuation-guided repair bouts between visits sustains continual learning across the stream, whereas wake-only continual-learning performance declines (Fig.~\ref{fig:repair}b). We quantify the continual-learning performance using a score $\mathcal Q$ that balances acquisition and retention. It gives weight $1/2$ to success on the current task and $1/4$ to success on the task from each of the two preceding visits. Recovering trainability during an offline phase has a functional parallel in human sleep. Fattinger et al.\ used sounds to disrupt slow waves over the motor cortex during sleep. The next morning, the same area showed less recovery of its ability to change in response to new finger-movement training~\cite{Fattinger2017}.

We also find that whether repair helps or hurts depends on the novelty of the waking experience (Fig.~\ref{fig:repair}c). We vary the fraction of visits that introduce a new task, which also changes repeat frequency and stream length (Supplemental Sec.~\ref{sm:streams}). % When the stream consists entirely of familiar tasks ($0\%$ novelty), offline repair reduces the mean continual-learning score, as restoring learning capacity can disrupt previously acquired responses without providing a corresponding benefit for learning new tasks. In contrast, for every tested stream containing new tasks, the mean paired gain $\Delta\mathcal Q=\overline{\mathcal Q}_{\mathrm{offline}}-\overline{\mathcal Q}_{\mathrm{no\ sleep}}$ is positive, with the largest tested mean gain near $40\%$ novelty. Overbars denote averages over each network's stream. These results highlight a competition between preserving previously learned responses and maintaining the capacity to acquire new ones. The dependence on waking experience also has parallels in animals. In rats, greater waking exploration predicted stronger slow-wave activity during subsequent sleep even when waking duration was held constant~\cite{Huber2007Exploration}. Similarly, exposure to a socially enriched environment increased subsequent sleep in flies~\cite{Donlea2009}.
For familiar tasks ($0\%$ novelty), offline repair lowers the mean continual-learning score. At every tested nonzero novelty, however, the mean paired gain $\Delta\mathcal Q=\overline{\mathcal Q}_{\mathrm{offline}}-\overline{\mathcal Q}_{\mathrm{no\ sleep}}$ is positive, with the largest tested mean gain near $40\%$ novelty. Overbars denote averages over each network's stream. This sensitivity to waking experience has parallels in animals. In rats, greater waking exploration predicted stronger slow-wave activity during subsequent sleep even when waking duration was held constant~\cite{Huber2007Exploration}. In flies, exposure to a socially enriched environment increased subsequent sleep~\cite{Donlea2009}.

Because repair alters previously learned responses while restoring learning capacity, its effectiveness depends on how strongly those adjustments are applied, or the dose (Fig.~\ref{fig:repair}d). We vary the prescribed repair dose $D=N_{\mathrm r}\varepsilon$ by adjusting the update strength $\varepsilon$ at fixed sample count $N_{\mathrm r}$; the baseline comparison in Fig.~\ref{fig:repair}b uses $D=1.6$. The actual stiffness changes also depend on the local readings. The highest mean score occurs at an intermediate tested dose. At the first repair bout, smaller doses restore less learning capacity, while larger doses retain fewer previously learned responses. The dominance measurements help navigate this trade-off. Some memory erosion is unavoidable here because our repair phase is driven purely by internal random forces, with no reference to the waking tasks. Unlike biological sleep, which also reactivates waking representations to consolidate memories~\cite{Rasch2013}, this minimal offline phase models only the restoration of plasticity.

Blind relaxation toward rest stiffness also restores learning capacity without these measurements. At matched total absolute change in log stiffness, summed over bonds and repair updates, guided repair achieves a higher continual-learning score in 49 of the 50 networks (Supplemental Fig.~\ref{fig:s-decay}). The measurement-guided repair rule therefore provides a better balance between acquisition and retention in this comparison.

Finally, withholding repair produces a rebound in subsequent repair demand, paralleling the compensatory sleep after deprivation used as a behavioral criterion for sleep in animals~\cite{Campbell1984,Shaw2003}. In a separate 12-visit stream, regular repair keeps sleep pressure $P$ within a bounded range: pressure rises during wake and drops after each repair bout. Skipping the final three or six repair bouts allows pressure to accumulate (Fig.~\ref{fig:repair}e). When repair resumes, deprived networks require a larger recovery dose $D$ to return to the same imposed target learning capacity, $\cpty/B\geq0.75$ (Fig.~\ref{fig:repair}f). Across the 50 paired networks, the mean extra dose required compensates for only part of the skipped repair: $75.3\%$ of the skipped dose after three omitted bouts and $52.2\%$ after six. This mirrors the partial recovery of lost sleep observed after biological sleep deprivation~\cite{Shaw2003}.

\subsection{Task activity biases the repair signal}
\label{sec:separation}
\label{sec:alternation}

\begin{figure}[!tb]
  \centering
  \includegraphics[width=\columnwidth]{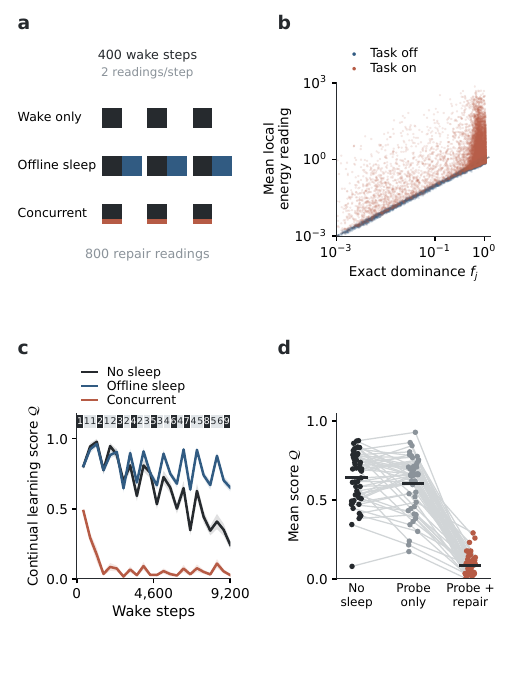}
  \caption{\textbf{Task activity biases the repair signal, and offline repair sustains higher performance.}
(a)~Wake-only, offline, and concurrent schedules with matched reading counts; block widths represent measurement counts.
(b)~Task-off and task-on energy readings versus exact dominance for trainable bonds across all 50 networks, using 200 paired fields per network; dashed line: unbiased reading.
(c)~Continual-learning score $\mathcal Q$ for the three protocols; means and standard errors over 50 paired networks on the standard stream.
(d)~Noise control in 50 matched networks. ``Probe only'' adds wake noise; ``probe plus repair'' also applies concurrent repair. Dots give each network's stream-averaged $\mathcal Q$, lines pair networks, and horizontal marks show means.}
  \label{fig:separation}
\end{figure}

The success of repair raises the question of why it should require a separate phase. A learner that could restore itself while acquiring a task would not need to interrupt its interaction with the environment. The obstacle identified in Sec.~\ref{theory:contamination} is that this interaction also enters the measurement used for repair. We test that prediction by applying the same fluctuation fields to each post-prelude network with and without task activity. Adding the task shifts the mean energy readings upward in all 50 networks (Fig.~\ref{fig:separation}b). The repair signal therefore depends on what the network is currently doing, even when its structure is held fixed.

This dependence makes the timing of repair consequential. We compare the schedules in Fig.~\ref{fig:separation}a at matched energy measurement counts and prescribed repair dose. Offline repair uses undriven readings after each visit; concurrent repair uses the two driven wake states. Fully concurrent repair produces much poorer continual-learning performance than offline repair (Fig.~\ref{fig:separation}c). The task-induced shift has the weakening bias derived in Theory: the bonds involved in carrying the task contribute extra energy to a signal intended to guide restoration. Moving the measurements offline removes this contribution. The comparison gives a specific physical reason for separating adaptation from repair, consistent with the biased-input rationale for offline renormalization in the synaptic homeostasis hypothesis~\cite{Tononi2014}.

The probe-only control tests whether the fluctuations themselves account for the failure of concurrent repair (Fig.~\ref{fig:separation}d). Adding the repair probe fluctuations without repair updates leaves mean performance close to that of learning alone, whereas adding concurrent repair lowers performance below both controls in every tested network. Probe noise alone, therefore, does not account for the collapse. Together with the task-on and task-off measurements, this supports the interpretation that the problem is repair acting on a signal contaminated by task activity. The next comparison uses the same noisy wake procedure across protocols.

\subsection{Separating strengthening and weakening}
\label{sec:sign}

\begin{figure}[!tb]
  \centering
  \includegraphics[width=\columnwidth]{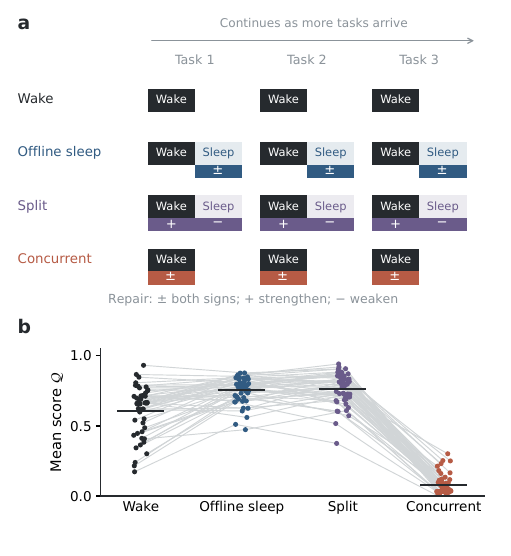}
  \caption{\textbf{Separating strengthening and weakening preserves continual-learning performance.}
(a)~Wake, offline, split, and concurrent schedules. Split permits positive repair during wake and negative repair offline; arrows indicate continued task presentations. Blocks are shown in phase order; widths are schematic.
(b)~Stream-averaged $\mathcal Q$ for 50 paired networks at matched repair measurement budgets. Dots identify networks, lines pair them, and bars mark means. All conditions use the same noisy wake procedure.}
  \label{fig:sign}
\end{figure}

Because task activity contributes positively to the measured bond energy, it introduces an asymmetry between strengthening and weakening repair. A bond identified as underdominant remains underdominant, whereas a bond identified as overdominant may simply be carrying a large task-induced strain. Recall that restoring the network's capacity to learn new tasks requires weakening overdominant bonds as well as strengthening underdominant ones. This asymmetry suggests that strengthening could occur during wake, while weakening should be reserved for an offline phase in which task activity no longer contaminates the measurement. We therefore test whether strengthening and weakening can be assigned to different phases. We implement the split rule motivated in Sec.~\ref{theory:contamination}: strengthening repair is allowed during wake, while weakening repair is reserved for the offline phase (Fig.~\ref{fig:sign}a). This allows part of repair to occur during learning while keeping weakening away from the task-contaminated signal.

We control for the noise and measurement procedures when comparing repair schedules. All conditions use the same noisy wake procedure. The three repair schedules use the same averaged energy estimator, total estimate count, and prescribed repair dose. Split repair assigns half its estimates to each phase; sign selection determines the stiffness changes actually applied (Supplemental Sec.~\ref{sm:schedules}). We find that split repair sustains a higher mean continual-learning score than learning alone or fully concurrent repair (Fig.~\ref{fig:sign}b). Its performance is close to that of fully offline repair, with no established advantage over it. Thus, repair does not have to be entirely offline to support continual learning: separating strengthening from weakening retains the benefit while allowing some restorative adjustments during wake. 
Importantly, reversing the sign assignment, so that weakening occurs during wake and strengthening offline, substantially reduces the continual-learning score in all 50 networks, despite using the same allocation of measurements between phases (Supplemental Fig.~\ref{fig:s-reverse}). Thus, the benefit of split repair depends not simply on separating restorative adjustments between wake and offline phases, but on which adjustments occur in each phase.

This separation parallels induced firing-rate homeostasis in rat visual cortex. Upward recovery after visual deprivation occurs during active wake, whereas downward recovery after eye reopening occurs during sleep~\cite{Hengen2016,TorradoPacheco2021}. Our network provides a possible physical explanation for why downward adjustment is restricted to sleep: task activity biases the signal used for repair. Whether the same explanation contributes to the biological separation remains a testable hypothesis.

\section{Discussion}
\label{sec:discussion}

We investigated whether the need for sleep requires biology or whether it can arise directly from the mechanics of continual learning. In a physical learner, adapting to new tasks changes internal degrees of freedom within finite structural limits. For any linear network governed by a positive quadratic form at fixed topology, boundary conditions, and rank, the conserved budget $\cpty+P=\budget$ [\cref{eq:budget}] makes this trade-off quantitative. Learning specializes the network by driving bond dominance toward its limits, and any increase in polarization $P$ reduces the learning capacity $\cpty$ available for future tasks by the same amount. Local random perturbations provide the measurements needed to guide repair, which recovers trainability with a better balance between acquisition and memory retention than a repair rule based on blind relaxation.

In this framework, the need to restore adaptability and the need to go offline have connected but distinct physical origins. Learning alters the network's structure, whereas ongoing task activity biases the measurements used to restore it. When task drive is present during probing, task strain adds extra energy to the bonds currently in use [\cref{eq:contam}], shifting the repair signal toward weakening active pathways. Separating learning and repair into distinct phases removes this interference and sustains continual learning. This physical mechanism also clarifies when restoration could remain online: if a learner could distinguish task-driven strain from probing fluctuations, or if sufficiently disengaged regions of an extended network underwent repair while active regions carried the task, restoration would not require global quiescence.

Even though we started from a purely nonbiological system, we found several functional parallels to biological sleep. This suggests that these phenomena reflect general physical constraints on continual adaptation. While biology implements sleep through specific biochemical pathways, any physical learner governed by a positive quadratic form faces the same budget, regardless of its substrate. Dominance polarization provides a mechanical analogue of synaptic saturation~\cite{Tononi2014,Kuhn2016}, while fluctuation-guided repair parallels the sleep-dependent recovery of cortical plasticity~\cite{Tononi2014,Fattinger2017}. Withholding repair allows polarization to accumulate and increases the subsequent repair dose needed to reach a target learning capacity, reminiscent of  the behavioral criterion of compensatory rebound after sleep deprivation~\cite{Campbell1984,Shaw2003,Borbely2022}. The benefit of repair also depends on the novelty of waking experience, echoing experience-dependent sleep need in animals~\cite{Huber2007Exploration,Donlea2009}. Finally, because task strain only inflates local energy readings ($\cont_b\geq0$), strengthening can remain effective during wake, whereas weakening benefits from being restricted to an offline phase, paralleling the separation of upward and downward firing-rate homeostasis between wake and sleep in rat visual cortex~\cite{Hengen2016,TorradoPacheco2021}.

At the same time, it is important to distinguish the functional need for an offline phase from its autonomous emergence. In our simulations, both the two-state wake updates and the switch between wake and repair are applied by an external supervisor, and the recovery target is externally specified. In living systems, these processes are regulated internally: organisms operate out of equilibrium, harvesting and dissipating energy to sense errors, modify their own structure, and regulate when they disengage from the environment. Embedding our mechanical learner and its conserved budget into an autonomous, out-of-equilibrium framework is a natural next step. Doing so will allow us to ask under what energetic conditions an offline repair phase emerges spontaneously without external scheduling, and how internal repair demand couples to energy cycles in the environment to produce circadian-like rhythms.

Beyond autonomous regulation, this physical setting offers a controlled way to study other features of biological sleep. While our repair protocol uses uncorrelated random bond forces, replacing white noise with structured excitations such as slow waves could reveal how different patterns of internal activity in REM- and non-REM-like states shape the trade-off between learning-capacity recovery and memory preservation. Extending the network from static mechanical allostery to time-dependent viscoelastic or inertial dynamics would likewise allow the system to learn temporal sequences rather than purely spatial input-output mappings. Another intriguing extension is to physical learners, such as granular packings and collections of living cells, in which learning occurs through structural rearrangements~\cite{guo2026learning,ameen2026training}. Could offline periods of reorganization restore their capacity to learn new tasks while preserving previously acquired memories? Finally, because unguided repair restores learning capacity without reference to prior tasks, it inevitably erodes stored memories. Motivated by evidence that sleep-like replay reduces forgetting~\cite{Rasch2013,Tadros2022}, introducing an offline phase that reactivates previously learned task drives during repair will allow us to test how a physical analogue of memory reactivation, or \emph{dreaming}, can protect recently learned tasks while restoring the capacity to learn.

\section*{Acknowledgements}

JMS acknowledges the National Science Foundation under DMR-2204312. NP acknowledges the National Science Foundation under DMR CAREER-2541744. This research was supported in part by grant NSF PHY-2309135 to the Kavli Institute for Theoretical Physics (KITP).

\section*{Competing interests}
The authors declare no competing interests.

\bibliography{refs_sleep}

\clearpage
\flushbottom
\AddToHookNext{shipout/after}{\setcounter{page}{0}\gdef\thepage{S\arabic{page}}}
\appendix
\makeatletter
\@removefromreset{equation}{section}
\makeatother
\section*{Supplemental Material}
\phantomsection\label{supp:start}
\setcounter{equation}{0}
\renewcommand{\theequation}{S\arabic{equation}}
\renewcommand{\theHequation}{supp.\arabic{equation}}
\renewcommand{\thefigure}{S\arabic{figure}}
\renewcommand{\theHfigure}{supp.\arabic{figure}}
\setcounter{table}{0}
\renewcommand{\thetable}{S\arabic{table}}
\renewcommand{\theHtable}{supp.\arabic{table}}
\input{supplementary.tex}
\end{document}

%% file: supplementary.tex
% Supporting derivations and figure placements.
\setcounter{figure}{0}
\makeatletter
\newenvironment{suppfigure}{\par\addvspace{8pt}\noindent\begin{minipage}{\columnwidth}\def\@captype{figure}\centering}{\end{minipage}\par\addvspace{8pt}}
\makeatother

\section{Supporting derivations}
\label{supp:theory}

\subsection{Bond dominance and the conserved budget}
\label{app:cavity}
\label{app:budget}
This section derives the cavity form of the dominance [Eq.~\eqref{eq:loadfrac}] and the conserved budget $\cpty+P=\budget$ [Eq.~\eqref{eq:budget}]. Let $\Hmat_{-b}=\Hmat-k_b\vect g_b\vect g_b^{\mathsf T}$ be the Hessian with bond $b$ removed. If removing bond $b$ preserves rank on the allowed node subspace, the Sherman--Morrison identity gives
\begin{equation}
  \Hmat^{+} = \Hmat^{+}_{-b}
  - \frac{k_b\, \Hmat^{+}_{-b}\, \vect{g}_b \vect{g}_b^{\mathsf{T}}\, \Hmat^{+}_{-b}}
         {1 + k_b\, \vect{g}_b^{\mathsf{T}} \Hmat^{+}_{-b} \vect{g}_b} .
  \label{eq:app-hinv}
\end{equation}
Contracting with $k_b\vect g_b$ gives $f_b=S_b/(1+S_b)\in[0,1)$. If removal lowers the rank, the projector definition gives $f_b=1$ directly and the finite-$S_b$ cavity expression does not apply. Summing the projector diagonal yields
\begin{equation}
  \sum_b f_b = \Tr\!\Big(\Hmat^{+} \sum_b k_b\, \vect{g}_b \vect{g}_b^{\mathsf{T}}\Big) = \Tr\!\left(\Hmat^{+}\Hmat\right) = n .
  \label{eq:app-sum}
\end{equation}
With $\bar f=n/N_b$ and $R=N_b-n$, this fixes the budget:
\begin{equation}
  \cpty + P = \sum_b f_b - 2\bar f \sum_b f_b + N_b \bar f^2 = n - \frac{n^2}{N_b} = \frac{nR}{N_b}.
  \label{eq:app-budget}
\end{equation}
Using $\delta\vect f=\Lmat\,\delta\lnk$ gives
\begin{equation}
  \dot{\cpty} = -\dot{P} = -2\, \left(\vect{f}-\bar f\,\vect{1}\right)^{\mathsf{T}} \Lmat \, \frac{\dd \lnk}{\dd t} .
  \label{eq:cdot}
\end{equation}

\subsection{Dominance response and pressure descent}
\label{app:laplacian}
This section derives the dominance response [Eq.~\eqref{eq:laplacian}] and the pressure descent [Eq.~\eqref{eq:repair-descent}]. Both follow from the potential $\Phi(\lnk)=\ln\pdet\Hmat$, where $\pdet\Hmat$ is the pseudo-determinant, the product of the nonzero eigenvalues of $\Hmat$; its gradient is $\vect f$ and its Hessian is $\Lmat$. Because the zero modes of $\Hmat$ do not change while all stiffnesses are positive, Jacobi's formula gives
\begin{equation}
  \begin{aligned}
  \frac{\partial \Phi}{\partial \ln k_b} &= k_b\, \vect{g}_b^{\mathsf{T}}\Hmat^{+}\vect{g}_b = f_b ,\\
  \frac{\partial f_b}{\partial \ln k_c} &= \delta_{bc}\, f_b - \Amat_{bc}^2 = \Lmat_{bc} .
  \end{aligned}
  \label{eq:app-grad}
\end{equation}
Idempotency of $\Amat$ implies $\sum_{c\ne b}\Amat_{bc}^2=f_b(1-f_b)$. Thus $\Lmat$ has zero row sums and nonpositive off-diagonal entries, forming a positive semidefinite weighted graph Laplacian. Writing $\vect q=\vect f-\bar f\vect1$, the pressure gradient is $\nabla_{\lnk}P=2\Lmat\vect q$. While $\dd\lnk/\dd t=-\varepsilon\vect q$ is not Euclidean gradient descent, substituting it into Eq.~\eqref{eq:cdot} establishes Eq.~\eqref{eq:repair-descent} when all bonds are adjustable.

\begin{suppfigure}
  \centering
  \includegraphics[width=\columnwidth]{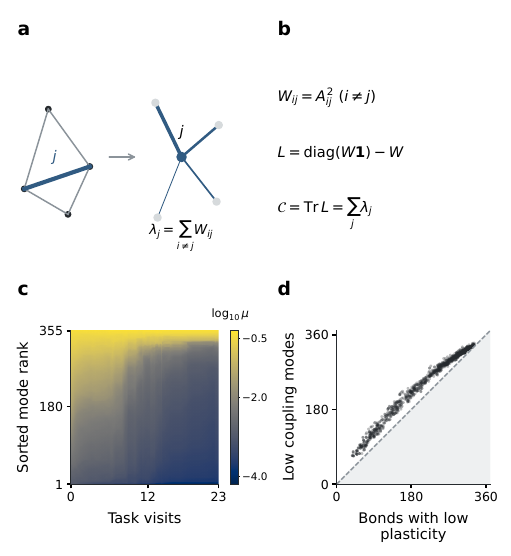}
  \caption{\textbf{The bond-coupling graph and its spectrum.}
(a,b)~Mapping from springs to coupling-graph nodes with edge weights $\Amat_{ij}^2$ and degree $f_j(1-f_j)$.
(c)~Spectra for one realization across the 23-visit stream after prelude tasks 1--3, including the uniform zero mode; colors show $\log_{10}\mu$ clipped to $[-4.2,-0.3]$.
(d)~Counts of low-plasticity bonds ($\lambda<0.01$) and low-eigenvalue modes ($\mu<0.02$) at 1150 endpoints across 50 networks. The excluded region follows from Sec.~\ref{sm:statistics}.}
  \label{fig:s-coupling}
\end{suppfigure}

\subsection{Wake response and the plasticity bound}
\label{app:wake}
\label{app:landing}
Under nodal load $\vect F$, the rank-preserving cavity construction gives
\begin{equation}
  t_b = k_b\, \psi_b\,\left(1 - f_b\right),
  \qquad
  \psi_b \equiv \vect{g}_b^{\mathsf{T}}\, \Hmat^{+}_{-b}\, \vect{F} .
  \label{eq:app-tension}
\end{equation}
To first order in $\Delta k_b/k_b$, Eq.~\eqref{eq:mml} becomes
\begin{equation}
  \frac{\dd \ln k_b}{\dd t}
  = \gamma\, k_b\, \tilde{P}_b \left(1 - f_b\right)^2 ,
  \qquad
    \tilde{P}_b \equiv -\, \psi^{\mathrm{FF}}_b\, \psi^{\mathrm{FB}}_b .
  \label{eq:app-mmldyn}
\end{equation}
For a trainable bond $b\neq p,q$, the task response obeys
\begin{equation}
  \begin{aligned}
  \delta \eta_\alpha &= \nabla_{\lnk}\eta_\alpha \cdot \delta \lnk^{\mathrm{sleep}},\\
  \frac{\partial \eta_\alpha}{\partial \ln k_b}
  &= -\sqrt{k_q/k_p}\;\Amat_{qb}\Amat_{bp} .
  \end{aligned}
  \label{eq:cost}
\end{equation}
A finite change of bond $b$ to bond dominance $f^\star$ gives
\begin{equation}
  \Delta \Amat_{qp} = -\,\frac{f^\star - f_b}{f_b(1-f_b)}\;\Amat_{qb}\Amat_{bp},
  \qquad
  \left|\Delta \Amat_{qp}\right| \le \tfrac12 \left| f^\star - f_b \right| .
  \label{eq:landing}
\end{equation}
Specifically, shifting $\ln k_b$ by $\Delta$ in Eq.~\eqref{eq:app-hinv} yields
\begin{equation}
  \Delta \Amat_{qp} = -\frac{(e^{\Delta}-1)\,\Amat_{qb}\Amat_{bp}}{1 + (e^{\Delta}-1) f_b},
  \qquad
  f^\star = \frac{e^{\Delta} f_b}{1 + (e^{\Delta}-1) f_b},
  \label{eq:app-landing}
\end{equation}
so $\ln[f^\star/(1-f^\star)]=\ln[f_b/(1-f_b)]+\Delta$. Eliminating $\Delta$ yields Eq.~\eqref{eq:landing}. Taking $\Delta\to0$ recovers the response derivative, while the row-sum identity gives
\begin{equation}
  \left|\Amat_{qb}\Amat_{bp}\right| \le \tfrac12\left(\Amat_{qb}^2 + \Amat_{bp}^2\right) \le \tfrac12 f_b(1-f_b) .
  \label{eq:app-bound}
\end{equation}

\subsection{Local fluctuations and task contamination}
\label{app:probe}
\label{app:contamination}
For bond forces $\vect F=(\vartheta\Kmat)^{1/2}\vect\xi$ with unit covariance, mechanical equilibrium $\vect u=\Hmat^+\Gmat^{\mathsf T}\vect F$ gives
\begin{equation}
  \left\langle \vect{u}\vect{u}^{\mathsf{T}} \right\rangle
  = \Hmat^{+}\Gmat^{\mathsf{T}} (\vartheta \Kmat)\, \Gmat\, \Hmat^{+}
  = \vartheta\, \Hmat^{+} ,
  \qquad
  \left\langle e_b^2 \right\rangle = \vartheta\, \vect{g}_b^{\mathsf{T}}\Hmat^{+}\vect{g}_b .
  \label{eq:app-equip}
\end{equation}
Multiplying by $k_b/\vartheta$ yields Eq.~\eqref{eq:equip}, and averaging Eq.~\eqref{eq:creep} over fast fluctuations recovers Eq.~\eqref{eq:creepmean}. During wake, total extension is $e_b=e_b^{\mathrm{task}}+e_b^{\mathrm{probe}}$. For independent, zero-mean probe noise,
\begin{equation}
  \left\langle k_b e_b^2 \right\rangle = k_b \left(e_b^{\mathrm{task}}\right)^2 + 2 k_b e_b^{\mathrm{task}} \left\langle e_b^{\mathrm{probe}}\right\rangle + \vartheta f_b = \vartheta\left(f_b + \cont_b\right).
  \label{eq:app-contam}
\end{equation}

\begin{suppfigure}
  \centering
  \includegraphics[width=\columnwidth]{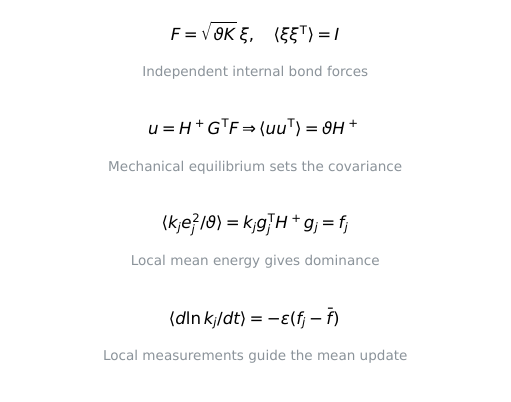}
  \caption{\textbf{Local bond noise gives the dominance estimator.}
Covariance and mean repair derivations follow Sec.~\ref{app:probe}; finite updates and port constraints follow Secs.~\ref{sm:schedules} and \ref{sm:network}.}
  \label{fig:s-probe}
\end{suppfigure}

\subsection{Dose near the rest state}
\label{app:dose}
Linearizing Eq.~\eqref{eq:creepmean} about $f_b=\bar f$ and keeping only the diagonal of $\Lmat$ gives
\begin{equation}
  \frac{\dd}{\dd t}\left(f_b - \bar f\right) \simeq -\varepsilon\, \bar f(1-\bar f)\left(f_b - \bar f\right).
  \label{eq:app-dose}
\end{equation}
A bout of nominal dose $D=\int_0^T\varepsilon(t)\,\dd t$ preserves a fraction
\begin{equation}
  s \simeq \exp\!\left[ -D\, \bar f (1-\bar f) \right]
  \label{eq:dose}
\end{equation}
of each dominance mismatch. This diagonal approximation estimates local relaxation, not task retention.

\subsection{Checks on the budget}
\label{app:pr}
\label{app:series}
From $\sum_b f_b=n$ and $0\leq f_b\leq1$, the dominance participation ratio satisfies
\begin{equation}
  \mathrm{PR}(\vect{f}) = \frac{\left(\sum_b f_b\right)^2}{\sum_b f_b^2}
                        = \frac{n^2}{\sum_b f_b^2},
  \qquad
  n \le \mathrm{PR}(\vect{f}) \le N_b .
  \label{eq:app-pr}
\end{equation}
For two springs in series between fixed walls ($\Gmat=(1,-1)^{\mathsf T}$, $\Hmat=k_1+k_2$, $n=R=1$),
\begin{equation}
  f_1 = \frac{k_1}{k_1+k_2},
  \qquad
  \Amat = \begin{pmatrix} f_1 & -\sqrt{f_1 f_2} \\ -\sqrt{f_1 f_2} & f_2 \end{pmatrix},
  \qquad
  \cpty + P = \tfrac12 .
  \label{eq:app-series}
\end{equation}

\section{Numerical methods}
\label{sm:simulation}
\input{supplementary_methods.tex}

%% file: supplementary_methods.tex
\subsection{Networks and task streams}
\label{sm:network}
\label{sm:streams}
We use 50 periodic networks with 128 nodes and $354$--$366$ bonds in a square box of side $L$, generated from relaxed bidisperse disk packings (radius ratio $1.4$, packing fraction $0.84$) by connecting nodes within $1.3$ times the sum of their radii. Geometry, rest lengths, and connectivity are fixed. Each network has 12 disjoint input--output bond pairs with midpoint separation $(0.30\pm0.05)L$, and all port bonds have fixed stiffness $k_0=1$. Pinning one node removes translations ($n=254$), so each network has its own $\bar f=n/N_b$ and $B=n-n^2/N_b$.

All protocols branch from identical network states after a wake-only prelude of three tasks ($480$ steps each). The prelude gives every protocol the same already-trained starting state, so retention can be scored from the first stream visit: at visit 1, the score $\score_v$ uses the last two prelude tasks (Sec.~\ref{sm:statistics}), and the $\nu=0$ stream repeats the prelude tasks. The standard stream has 23 visits of $400$ wake steps, with 9 new tasks spaced across 14 cyclic repeats. The novelty $\nu$, the fraction of visits introducing a new task, takes the values $(0,0.2,9/23,9/14,1)$ with $(45,45,23,14,9)$ visits; nonzero $\nu$ introduces the same 9 tasks, and $\nu=0$ repeats the prelude tasks. In the presented-first control, each of the 9 new tasks is trained independently for 400 steps from the post-prelude state ($450$ runs; Fig.~\ref{fig:setup}d).

\subsection{Dimensionless parameters}
\label{sm:units}
Stiffnesses are measured in units of the port stiffness $k_0$ and forces in units of the drive $\tau$, so the dimensionless learning rate is $\gamma\tau^2/k_0=100$ (the code uses $\tau=0.1$ and $\gamma=10^4$). In these units, the update in Eq.~\eqref{eq:mml} gives the relative change $\Delta k_j/k_j=-(\gamma\tau^2/k_0)\,\hat t_j^{\mathrm{FF}}\hat t_j^{\mathrm{FB}}/\hat k_j$, with $\hat t=t/\tau$ and $\hat k=k/k_0$. Most bonds carry only a small part of the input load: at the first wake step of the stream, $|\hat t_j^{\mathrm{FF}}\hat t_j^{\mathrm{FB}}|$ has median $6\times10^{-5}$ and maximum $7\times10^{-3}$. The resulting relative stiffness changes are small (Fig.~\ref{fig:s-step}): the median bond changes by $0.09\%$, 90\% of bonds by less than $1.4\%$, and the largest by $16\%$, while the output error decreases by about $1\%$ per step. This value of $\gamma$ therefore lets a task be learned within a 400-step visit while keeping each update small.

\begin{suppfigure}
  \centering
  \includegraphics[width=\columnwidth]{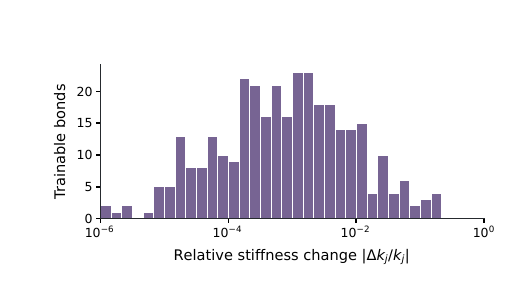}
  \caption{\textbf{Relative stiffness changes in one wake step.}
Distribution of $|\Delta k_j/k_j|$ over trainable bonds at the first wake step of the stream in one network.}
  \label{fig:s-step}
\end{suppfigure}

\subsection{Fluctuation measurements}
\label{sm:probes}
Independent standard normal variables $\xi_b$ on all bonds (including ports) generate forces
\begin{equation}
  F_b = \sqrt{\vartheta\, k_b}\;\xi_b ,
  \qquad
  \langle \xi_b \xi_c \rangle = \delta_{bc} ,
  \label{eq:probe}
\end{equation}
applying nodal load $\Gmat^{\mathsf T}\vect F$. Trainable bonds read $\widehat f_j = k_je_j^2/\vartheta$ without mean subtraction. Probe intensity is calibrated once per network from the post-prelude state and first new task:
\begin{equation}
 E_{\mathrm{ref}}=\frac12\sum_j k_j\big[(e_j^{\mathrm{FF}})^2+(e_j^{\mathrm{FB}})^2\big],
 \qquad \vartheta=\frac{rE_{\mathrm{ref}}}{n}.
 \label{eq:noise-calibration}
\end{equation}
Undriven readings do not depend on $\vartheta$, because extensions scale as $\sqrt{\vartheta}$, so $r=1$ for undriven single-field repair is a convention. Noisy wake and batch-averaged estimates use $r=0.1$, which keeps the probe weak enough that probe noise alone leaves performance close to learning alone (Fig.~\ref{fig:separation}d). In noisy wake, independent probe fields act on the feedforward and feedback states, and the output error is set by the noisy feedforward tension.

Figure~\ref{fig:repair}a compares running averages of undriven readings at fixed stiffness with the exact dominances $\Amat_{bb}$. Figure~\ref{fig:separation}b compares 200 paired task-off and task-on fields per network at fixed post-prelude stiffnesses.

\subsection{Repair schedules}
\label{sm:schedules}
\label{sm:updates}
Each repair update sets $k_j\leftarrow k_j\exp(g_j)$ on trainable bonds, with $g_j=\varepsilon(\bar f-\widehat f_j)$ clipped to $[-4,4]$; the clip is a safeguard that no reported run reaches. Wake repair uses the stiffnesses from before the learning step and is applied after it.

The baseline comparison (Figs.~\ref{fig:repair}--\ref{fig:separation}) uses $\varepsilon=0.002$, so $D=800\varepsilon=1.6$, within $0.03$ in mean $\score$ of the best tested dose ($D=0.8$; Fig.~\ref{fig:repair}d). Offline repair uses 800 undriven steps after each visit, and concurrent repair uses two noisy-wake readings per wake step, so both take 800 readings per visit. Dose sweeps vary $\varepsilon=D/800$ over $D\in\{0.2,0.4,0.8,1.6,3.2,6.4\}$. The probe-only control applies wake noise without repair, using independent random streams.

In the split comparison (Fig.~\ref{fig:sign}), all arms share the same noisy wake, and each estimate averages 64 fresh fields ($800$ estimates, or $51{,}200$ fields, per visit). Concurrent repair uses two estimates per wake step, and offline repair uses 800 estimates after each visit. Split repair applies two positive-only updates on every other wake step ($400$ in total) and 400 negative-only updates offline, after averaging and capping.

\subsection{Readouts and statistics}
\label{sm:statistics}
Noise-free responses are evaluated at $t_m=0,50,\ldots,400$:
\begin{equation}
  h_\alpha(t) = \Theta\!\left(\etol - \left|\eta_\alpha(t) - \eta^\star\right|\right),
  \label{eq:held}
\end{equation}
\begin{align}
  \score_v^{\mathrm{cur}} &= \frac{1}{9}\sum_{m=1}^{9} h_{\alpha_v}(t_m),
  \label{eq:scorecur}\\
  \score_v &= \frac{1}{9}\sum_{m=1}^{9}\left[\tfrac12 h_{\alpha_v} + \tfrac14 h_{\alpha_{v-1}} + \tfrac14 h_{\alpha_{v-2}}\right](t_m).
  \label{eq:scoremem}
\end{align}
The weights split $\score_v$ equally between acquiring the current task and retaining the two preceding ones. At visit 1, $\alpha_{v-1}$ and $\alpha_{v-2}$ are the last two prelude tasks. Because $\score$ includes the tasks of the two preceding visits, the prelude must contain at least two tasks for the first visits of every stream to be scored in the same way; we use three. Scores are averaged within each network before averaging over networks, and bootstrap intervals resample whole networks.

Bond dominances are computed from the singular vectors of $\Kmat^{1/2}\Gmat$, including the fixed ports. Figure~\ref{fig:resource}b plots the 354 nonzero eigenvalues $\mu$ of $\Lmat$ for the network of Fig.~\ref{fig:setup}b, which has 355 bonds; the uniform zero mode is omitted. With bond plasticity $\lambda_j=f_j(1-f_j)$, the bound $\Lmat\preceq2\operatorname{diag}(\lambda)$ implies $\mu_{(k)}\leq2\lambda_{(k)}$, which excludes the region below the diagonal in Fig.~\ref{fig:s-coupling}d. Figure~\ref{fig:s-soft} projects held task responses onto the eigenbases of the initial Hessian $\Hmat_0$, the initial Laplacian $\Lmat_0$, and the trained Hessian $\Hmat$ across the 293 held endpoint states.

\begin{suppfigure}
  \centering
  \includegraphics[width=\columnwidth]{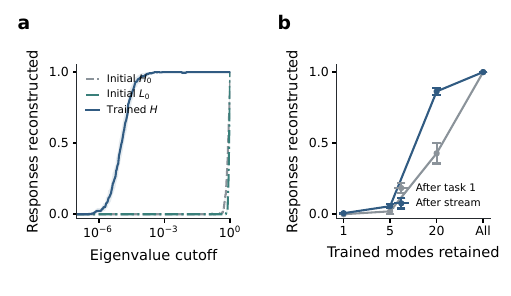}
  \caption{\textbf{Task-response reconstruction in fixed and trained bases.}
(a)~Reconstruction versus normalized eigenvalue cutoff in $\Hmat_0$, $\Lmat_0$, and trained $\Hmat$ (bands: 95\% network-bootstrap intervals).
(b)~Reconstruction from the lowest 1, 5, 20, or all 254 modes of $\Hmat$ (bars: mean $\pm$ s.e.m.\ over contributing networks; Sec.~\ref{sm:statistics}).}
  \label{fig:s-soft}
\end{suppfigure}

\section{Protocols and controls}
\label{sm:controls}

\subsection{Repair deprivation and recovery}
\label{sm:rebound}
In a 12-visit stream, the final 0, 3, or 6 offline repair bouts are omitted, with wake training and random seeds paired (Fig.~\ref{fig:repair}e). Deprived networks then receive up to 16,000 undriven readings ($D=32$), and the recovery dose is recorded when $\cpty/\budget$ first reaches $0.75$ ($P/\budget\leq0.25$), checked every $\Delta D=0.2$ (Fig.~\ref{fig:repair}f). The ordering of the arms is unchanged for $P/\budget$ thresholds of $0.20$ and $0.30$.

\subsection{Blind decay and expenditure matching}
\label{sm:decay}
Blind decay updates trainable bonds as $x_j\leftarrow(1-a)x_j$ with $x_j=\ln(k_j/k_0)$. To match the total stiffness expenditure of guided repair, $A_v=\sum_{\ell,j}|\Delta\ln k_j^{(\ell)}|$, wake and offline decay use step budgets $s_v=A_v/400$ and $s_v=A_v/800$, respectively, with $a=\min[1,s_v/\sum_j|x_j|]$. A reciprocal control halts guided repair once it reaches the realized expenditure of offline decay. At the reference budget, guided repair outperforms both wake and offline decay in all 50 networks (Fig.~S5a). Halted at offline decay's realized expenditure, it still outperforms offline decay in 49 of 50 (paired $\Delta\score=0.359$, 95\% CI $[0.330,0.383]$, 20,000 resamples;  Fig.~\ref{fig:s-decay}).

\begin{suppfigure}
  \centering
  \includegraphics[width=\columnwidth]{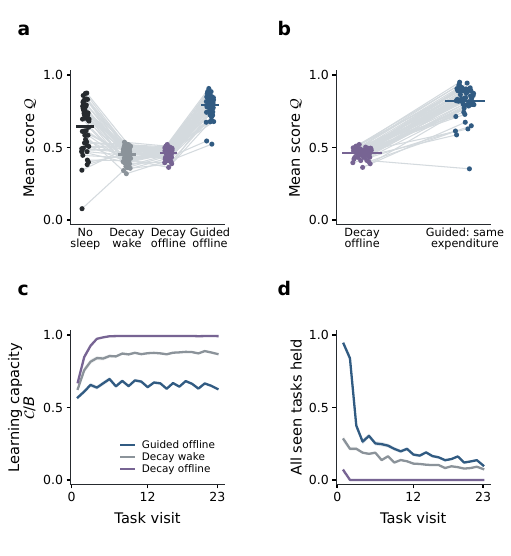}
  \caption{\textbf{Guided repair versus blind decay.}
(a)~Stream-averaged $\mathcal Q$ for reference-budget decay and guided repair.
(b)~Guided repair halted at offline decay's realized expenditure.
(c,d)~Post-visit learning capacity and fraction of encountered tasks held. Dots and lines pair 50 networks; bars and bands show mean $\pm$ s.e.m.\ (Sec.~\ref{sm:decay}).}
  \label{fig:s-decay}
\end{suppfigure}

\subsection{Reverse split}
\label{sm:reverse}
The reverse split exchanges the sign assignment of Sec.~\ref{sm:schedules}, applying negative updates during wake (400 estimates) and positive updates offline (400 estimates). The forward split outperforms the reverse split in all 50 networks (paired $\Delta\mathcal Q = 0.373$, 95\% CI $[0.329,0.421]$, 20,000 resamples; Fig.~\ref{fig:s-reverse}).

\begin{suppfigure}
  \centering
  \includegraphics[width=\columnwidth]{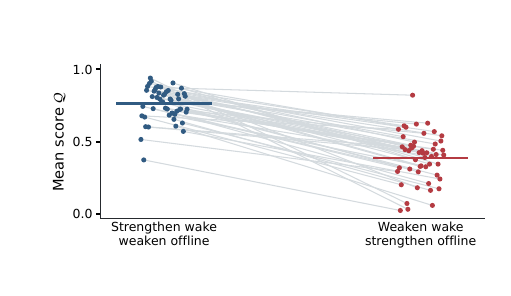}
  \caption{\textbf{Forward and reverse split.}
Stream-averaged $\mathcal Q$ at matched phase allocation and measurement counts (Sec.~\ref{sm:reverse}). Dots and lines pair 50 networks; bars mark means.}
  \label{fig:s-reverse}
\end{suppfigure}

\subsection{Arrival-position control}
\label{sm:arrival}
A linear probability model regresses first-visit acquisition of the 9 new tasks under the wake-only and offline protocols ($900$ outcomes) on arrival capacity $\mathcal C/B$, with fixed effects for network, arrival index, and protocol. Arrival capacity retains a positive slope of $1.19$ (95\% CI $[1.02,1.37]$, 2,000 network resamples; Fig.~\ref{fig:s-arrival}).

\begin{suppfigure}
  \centering
  \includegraphics[width=\columnwidth]{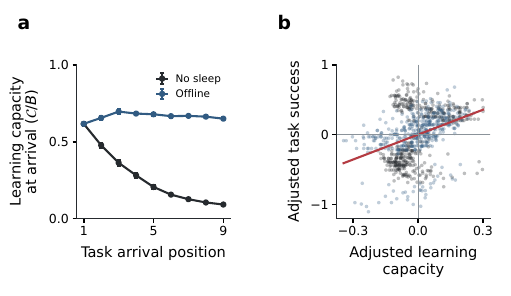}
  \caption{\textbf{Controlling for task arrival order.}
(a)~Learning capacity at matched task positions (mean $\pm$ s.e.m., 50 networks).
(b)~Residual acquisition versus residual capacity after removing network, position, and protocol fixed effects (Sec.~\ref{sm:arrival}).}
  \label{fig:s-arrival}
\end{suppfigure}